\documentclass[useAMS,usenatbib]{mn2e}
\usepackage{graphicx}

\title[Relativistic hybrid stars with super-strong magnetic fields]
{Relativistic hybrid stars with super-strong toroidal magnetic fields: An evolutionary track with QCD phase transition}
\author[N. Yasutake, K. Kiuchi, and K. Kotake]{Nobutoshi Yasutake$^1$\thanks{E-mail: yasutake@th.nao.ac.jp},
Kenta Kiuchi$^2$\thanks{kiuchi@gravity.phys.waseda.ac.jp},
Kei Kotake$^{1,3}$\thanks{kei.kotake@nao.ac.jp} \\
$^1$Division of Theoretical Astronomy, National Astronomical Observatory of Japan, 2-21-1 Osawa, Mitaka, Tokyo 181-8588, Japan~ \\
$^2$Department of Physics, Waseda University, 3-4-1 Okubo, Shinjuku-ku, Tokyo 169-8555, Japan~ \\
$^3$Center for Computational Astrophysics, National Astronomical Observatory of Japan, 2-21-1, Osawa, Mitaka, Tokyo, 181-8588, Japan~
}

\begin{document}

\date{Typeset \today ;  Accepted}

\pagerange{\pageref{firstpage}--\pageref{lastpage}} \pubyear{2008}

\maketitle

\label{firstpage}

\begin{abstract}
We investigate structures of hybrid stars, which feature quark core surrounded by a hadronic matter mantle, with  super-strong toroidal magnetic fields in full general relativity. Modeling the equation of state (EOS) with a first order transition by bridging the MIT bag model for the description of quark matter and the nuclear EOS by Shen et al., we numerically construct thousands of the equilibrium configurations to study the effects of the phase transition. It is found that the appearance of the quark phase can affect distributions of the magnetic fields inside the hybrid stars, making the maximum field strength about up to $30$ \% larger than for the normal neutron  stars. Using the equilibrium configurations, we explore the possible evolutionary paths to the formation of hybrid stars due to the spin-down of magnetized rotating neutron stars. We find that the energy release by the phase transition to the hybrid stars is quite large~($\la 10^{52}~\rm erg$) even for super strongly magnetized compact stars. Our results suggest that the strong gravitational-wave emission and the sudden spin-up signature could be observable signals of the QCD phase transition, possibly for a source out to Megaparsec distances.
\end{abstract}


\section{Introduction}
A very hot issue in hadronic and nuclear physics is to search 
the phase transition from baryons to their constitutes - deconfined quarks. 
Heavy ion colliders such as RHIC (Brookhaven) and LHC (CERN) are 
now on line 
to explore the QCD phase diagram for the high temperature and small 
baryon density regimes, for which 
lattice QCD calculations predict a smooth crossover to the QCD phase transition  
(see \citet{stephanov04} for review).

Conversely for the low temperature and high baryon 
density regimes, compact stars are expected to provide a unique window on
 the phase transition at their extreme density with super-strong magnetic field. 
It has been suggested long ago 
that quark matter may exist in the interior of compact objects 
\citep{itoh70,bodmer71,witten84}. 
Hybrid stars (and strange quark stars) 
are considered to be such objects, which feature quark cores surrounded 
by a hadronic matter mantle (or quark cores only) 
(for reviews, e.g., \citet{weber,glendenning01}).
Even if the relevant conditions could be reached in a laboratory 
in the near future \citep{senger}, the conditions prevailing in the compact 
stars are different from those produced in accelerators, i.e., 
the matter is 
long-lived, charge neutral and in $\beta$-equilibrium with respect to weak interactions.
It is therefore important to investigate the properties of such ``exotic'' stars, providing hints about the main features of matter at those extreme conditions. 

The formation of the quark cores in compact stars is expected 
to take place by a first-order phase transition \citep{glen92,glendenning01}. 
Albeit still very uncertain (e.g., \citet{horvath07}), such a transition would proceed 
by the conversion of initially metastable 
hadronic matter in the core into the new deconfined quark phase. The metastable phase 
could be formed as the central density of neutron stars exceeds a critical value, 
due to mass accretion, spin-down or cooling. Possible astrophysical 
cites are in the protoneutron stars during the collapse of 
of supernova cores (near the epoch of bounce \citep{takahara,yasutake07} 
or at the late postbounce phase \citep{gentile,nakazato08,sagert08})
 and in old neutron stars accreting from their companions \citep{benve,chau,lin06}. 
In whichever cases, the sudden nucleation of the exotic phase in the hadronic star 
will be accompanied by a core-quake and huge energy release of the gravitational binding 
energy. Such energy release has been proposed to explain the central engines of
 the gamma-ray bursts \citep{bombaci,bere}.  
Possible observables of these transient phenomenon should be 
 glitches, magnetar flares, and superbursts \citep{alford}. In addition,
 the detection
 of gravitational waves \citep{ioka03,yasutake07,lin06,abdikamalov08} and neutrinos 
\citep{nakazato08,sagert08} generated at the moment of the phase transition, 
should supply us implications to unveil the mechanism of the phase transition.

Here it should be noted that most of these calculations/estimations concerning 
the phase transition inside compact stars are limited to a non-rotating case, 
in which the Tolman-Oppenheimer-Volkoff equation is solved to obtain their structures \citep{haensel86,zdunik87,muto,drago04,yasutake09}. 
Exceptions are for \citet{gourgoulhon99,yasutake05, zdunik07}, in which 
relativistic equilibrium configurations of rotating strange stars, beyond the so-called 
slow rotation approximations (see references in \citet{glendenning01}), are 
constructed for estimating the energy release. 
In \citet{gourgoulhon99,yasutake05}, hadron matter is assumed to be converted fully 
to quark matter, leading to the 
formation of strange quark stars like in \citet{alcock86,benvenuto89,olesen91,lugones94}.
However, it is recently pointed out that the strange quark stars could be ruled out 
by their too fast spin-down rates via gravitational radiations from r-modes instability 
\citep{madsen}. The quasi-periodic oscillations (QPOs) of strange quark 
stars could not be reconciled with the observations \citep{anna}.
Moreover, \citet{mallick09} has recently 
 claimed that magnetars cannot convert to purely quark stars, 
but only to hybrid stars.
These suggest us to pay attention to hybrid stars rather than 
 stars made of purely deconfined quarks. 
In a very recent work by \citet{zdunik08}, 
the phase transition of rotating hybrid stars is discussed, however, 
 the equation of state (EOS) is made very idealistic, assumed to model a strong 
 phase transition, seemingly less sophisticated than the EOSs in the 
recent literature cited above.
In addition to rotation,  neutron stars observed in nature 
are magnetized with the typical magnetic field strength $\sim 10^{11}$--$10^{13}$ G
 \citep{Lyne}. The field strength is often much larger than the canonical value as $\sim
10^{15}$ G for a special class of the neutron stars such as magnetars
\citep{Lattimer:2007,woods}. In a series of our recent papers 
\citep{kiuchi08a,kiuchi08b,kiuchi08d}, we have studied equilibrium configurations 
 of relativistic magnetized compact stars to apply for 
the understanding of their formations and evolutions, but with hadronic EOSs.

Above situations motivate us to investigate the structures of relativistic
 hybrid stars with magnetic fields, and by using them, to estimate the energy 
release at the phase transition. This study is posed as an extension to the
 study by \citet{yasutake05}, in which the phase transition to 
the rotating strange stars was investigated. 
Due to the unavailability 
 of the method to construct a fully general relativistic star with arbitrarily 
magnetic structures (namely both with toroidal and poloidal fields),  
 we here consider the equilibrium with purely toroidal fields as in \citet{kiuchi08b}.
It is noted that the outcomes of the recent stellar evolution calculations 
\citep{Heger:2004qp} and the MHD(magnetohydrodynamics) simulations of core-collapse supernovae 
(\citet{kotake04,takiwaki04,obergaulinger06,luc2007,sawai2008,kiuchi08c,takiwaki09}), 
suggesting much dominance of the toroidal 
fields, are not in contradiction with the assumption.  
Following the scenario proposed in \citet{yasutake05}, 
 we consider 
the possible evolutionary tracks of a rapidly rotating and magnetized 
neutron star to a slowing rotating hybrid star due to the spin-down via 
gravitational radiation and/or magnetic braking. 
During the evolutions, the baryon mass
 and the magnetic field strength are taken to be constant for simplicity.
 The energy release can be estimated from the difference in the mass-energies 
between the hadronic star and the hybrid star along each sequence. By constructing
 thousands of equilibrium configurations, we hope to clarify 
the possible maximum energy release at the moment of the transition 
and discuss their astrophysical implications.

The paper is organized as follows. 
 The method for constructing the EOS with the phase transition 
and the numerical scheme for the stellar equilibrium configurations are briefly 
summarized in Sec.~\ref{sec2}.  Sec.~\ref{sec:result} is devoted to 
showing numerical results. Summary and discussion follow in Sec.~\ref{sec:summary}.
In this paper, we use geometrical units with $G=c=1$.

\section{Equation of State and Numerical Method}\label{sec2}
\subsection{Equations of State with a first-order phase transition}\label{sec:numsch}
As mentioned, we assume that the deconfinement of the quarks takes place 
at the first order phase transition.
In this case, a mixed phase can form and it is typically described using two 
separate EOSs, one for the hadronic and the other 
for the quark phase.
Here the bulk Gibbs construction is used to bridge the two phases. 
In the following, we 
briefly summarize the adopted EOSs for each phase and explain the features of 
the resulting EOS with the first order phase transition.

Since lattice QCD is yet to make solid predictions for large density regimes 
 in compact objects, the quark matter EOS is currently computed 
using phenomenological descriptions such as the MIT bag or the Nambu-Jona-Lasinio (NJL)
 models.
For the quark phase, we choose the EOS based on the very simple but 
widely applied MIT bag model (see \citet{weber,glendenning01} for reviews).
Using the model, one can express the energy density, $\epsilon$, and the pressure, 
$P$ of strange 
quark matter as functions of baryon number density, $n$ in the following form,
\begin{eqnarray}
\epsilon   &=& \sum_f \epsilon_f + B, \\
\epsilon_f &=& \frac{3m_f^4}{8 \pi^2} [x_f(2x_f^2+1)\sqrt{1+x_f^2}-\textrm{arsinh}~x_f],  \\
P           &=& n^2 \frac{d(\epsilon/n)}{d~n},
\end{eqnarray}
where $m_f$ is the quark mass of the flavor of $f$ taken to be
 $m_u=m_d=5$~MeV and $m_s=150$~MeV, $x_f=k_F^{(f)}/m_f$ is a normalized Fermi wave 
number of $k_F^{(f)}$, and $B$ is the bag constant.
Baryon number density of strange matter is $n = \frac{1}{3}
(n_u + n_d + n_s)$, where $n_f= k_ F^{(f)~3}/\pi^2 $ is the number density of $f$ 
quarks. We use a simple MIT bag model of self-bound strange quark 
matter \citep{chodos}, neglecting the quark interactions except 
for the confinement effects described by the bag constant  (see, e.g., references 
in \cite{luca}  for sophistication of the model). 
We set that the bag constants 
are 200~MeV fm$^{-3}$ and 250~MeV fm$^{-3}$ in this paper.  
These values seem consistent with the implications in the recent lattice QCD results 
\citep{ivanov05}. The quark EOS with lower bag constant than 200~MeV fm$^{-3}$ 
may be too soft to be compatible with the observation of the massive neutron star
 like pulsars Ter 5 I and J ($> 1.68 M_\odot$ with 95 \% confidence) \citep{ransom05,alford}. 
For the EOS of the hadron phase, we adopt the nuclear EOS developed by \citet{shen98}, 
which are often employed in recent MHD simulations relevant for magnetars
 (see \citet{Kotake:2006} for a review).
The Shen EOS is based on the relativistic 
mean field theory with a local density approximation, which has
 been constructed to reproduce the experimental data of masses and
radii of stable and unstable nuclei (see references in \citet{shen98}). 
At the maximum densities higher than two times of saturation density, muons may 
 appear \citep{wiringa88, akmal98}. However, we neglect it, 
since the muon contribution to pressure at the higher density has been 
pointed to be very small~\citep{douchin01}.  

In modeling the phase transition to quark matter, there is a main physical uncertainty, 
the critical density for the onset of the mixed phase. Under the MIT bag model, the transition is determined by the value of the bag constant. 
We use the bulk Gibbs construction to bridge the two phases using the technique 
developed by \citet{glen92}. In the transition, two conserved quantities are the 
baryon number density ($n_B$) and the electric charge density ($:Y_e n_B$ with $Y_e$ being the 
electron fraction), 
\begin{eqnarray}
n_B &=& \chi n_{B,Q}+(1-\chi) n_{B,H},\\
Y_e  n_B &=& \chi  Y_{C,Q} ~ n_{B,Q}+(1-\chi)~ Y_{C,H}~ n_{B,H}.
\end{eqnarray}
Here $\chi$ is the volume fraction of matter in the quark phase. The subscripts $H$ and 
$Q$ label the number density and the charge fraction, $Y_C$, 
in the hadronic and in the quark phase, respectively. Since matter in cold 
compact stars is in chemical 
equilibrium under $\beta$-decay with vanishing neutrino chemical potentials, 
the following equations have to be satisfied,
\begin{eqnarray}
\mu_e + \mu_p &=& \mu_n, \\
\mu_n &=& \mu_u+2 \mu_d, \\
\mu_p &=& 2 \mu_u+\mu_d,\\
\mu_d &=& \mu_s,  
\end{eqnarray}
together with the mechanical equilibrium, the equality of the pressure in the two phases,
\begin{eqnarray}
 P^Q = P^H.
\end{eqnarray}
Here for simplicity, the finite size effects on the phase transition~\citep{endo06, maruyama} are ignored. Five unknown variables to describe the mixed phase 
($\chi,  n_{B,Q}, n_{B,H}, Y_{C,Q},  
Y_{C,H}$) can be solved by the five sets of the equations of (3,4,6,7,9).

\begin{figure*}
\begin{center}
\includegraphics[width=17cm]{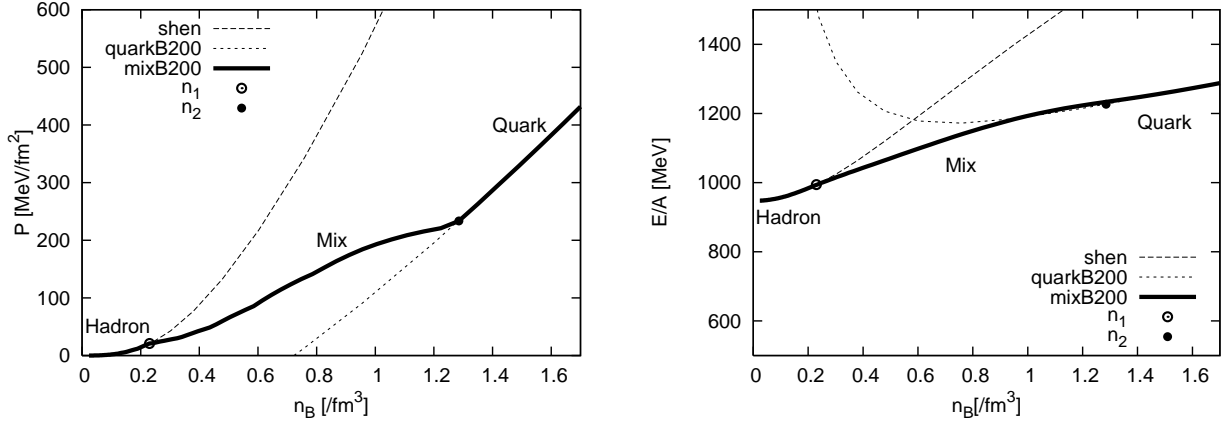} 
\caption{\label{fig:eos1} The constructed EOS under bulk Gibbs condition. 
The labels of ``quarkB200'' and ``mixB200'' indicate a pure quark matter EOS and an EOS with mixed phase with $B=200~{\rm MeV}~{\rm fm} ^{-3}$, whereas "Shen" indicates Shen EOS of pure hadronic matter. 
Left and right panels are the pressure and the energy per baryon as a function of the baryon number density, respectively. }
\end{center}
\end{figure*}

\begin{figure*}
\begin{center}
\includegraphics[width=17cm]{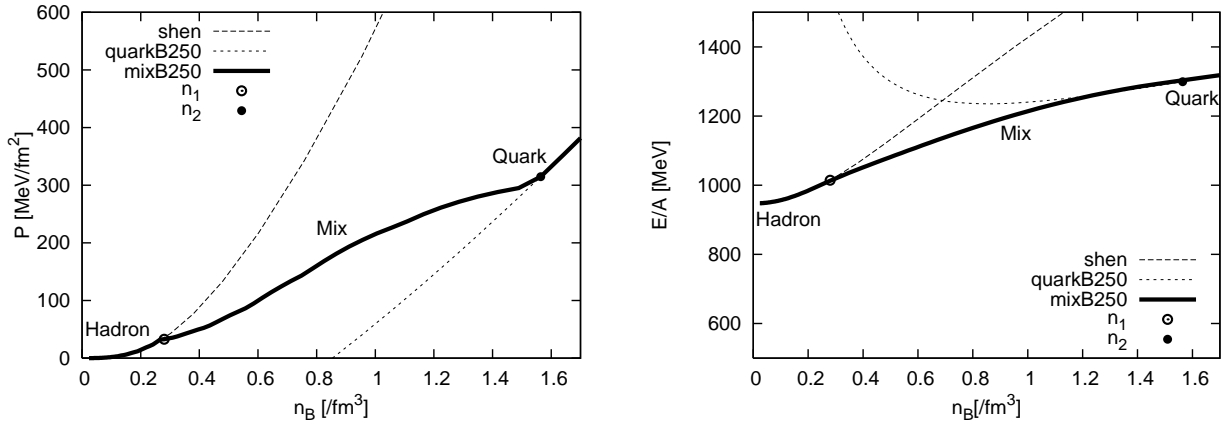} 
\caption{\label{fig:eos2} Same as Figure~1 but for $B=250~{\rm MeV}~{\rm fm} ^{-3}$. }
\end{center}
\end{figure*}

 Figures~\ref{fig:eos1} and \ref{fig:eos2} show the constructed EOS with the first-order phase transition.
For comparison, we also plot the hadronic ``Shen EOS'' mentioned above 
and pure quark EOS("quarkB200" and "quarkB250").
 The left and right panels are the pressure and the energy per baryon 
as a function of the baryon density, respectively. 
The labels of ``mixB200(quarkB200)'' and ``mixB250(quarkB250)'' indicate 
the two different 
bag constants, B=200~MeV fm$^{-3}$ and 250~MeV fm$^{-3}$.
The EOS consists of the three phases of the ``Hadron'', ``Mixed'', 
and the ``Quark''. 
The critical densities with $n_1$ (open circles) and $n_2$ (closed circles) 
marking the boundaries between the hadron, mixed, and quark phase, are also given in 
Table~\ref{tab:critical}. From the left panel and Table 1, it can be seen that the 
critical density for the mixed phase becomes higher for the larger bag constant.
 This is because for the large bag constant, the density should be larger to achieve the 
equilibrium between the mixed phase and the pure hadron phase,  remembering that the 
pressure of the quark matter becomes smaller with increasing the bag constant
(from Eqs. (1,2)). The higher $n_1$ for the large bag constant also leads to the 
higher $n_2$, the transition to the quark phase 
(Table 1). From the right panel of Figure~\ref{fig:eos1} and \ref{fig:eos2}, it is seen that the energy with the quark phase~("mixB200" and "mixB250") are clearly lower than the case only with the hadronic phase~("shen"), 
and also that the energy is generally lower for the smaller bag constant above $n_1$.

The actual conversion process from nuclear matter to quark matter has been of a topic 
of hot debate, such as detonation or deflagration \citep{drago05}.
 The range of the conversion timescale is estimated to be very wide, 
roughly $0.1-100$ sec, depending upon the bag constant, temperature, mass of neutron stars, and so forth ~\citep{olesen91}. Since our main aim of this paper is the estimation of the liberated energy by the conversion from neutron stars to hybrid stars, we do not consider the detailed combustion processes in this study.

Proto-neutron stars left after supernova explosion are very hot 
($\sim 50$ MeV) initially, however, cool down below $\sim 1 $MeV 
in some tens of seconds \citep{burrows86}. The newly formed neutron stars stabilize 
at practically zero temperature. As mentioned, in cold neutron stars, 
the $\beta$ equilibrium in the weak interactions can be well validated, with 
 neutrinos and antineutrinos freely escaping from the star. Combining the zero-temperature, zero-neutrino fraction, 
and the beta-equilibrium conditions with the charge neutrality condition, 
  the thermodynamic variables depending on the three parameters
 (e.g. the pressure as $P(\rho, Y_e, T)$ with $Y_e$ being the electron fraction), 
can only be determined by a single
variable (barotropic), which we take to be the energy density, 
namely $P(\epsilon)$ \citep{shapiro83} in the following.

In the following, we introduce the baryon mass density, $\rho_0$, for convenience,  defined to be the number density, $n$, multiplied a baryon mass, 
$m_0=1.6605 \times 10^{-24}$ g. 

\begin{table}
\centering
\begin{minipage}{70mm}
\caption{\label{tab:critical}
Critical densities making the transition to the mixed, $n_1$, and the quark phase, 
$n_2$, respectively. }
\begin{tabular}{ccc}
\hline\hline
EOS       & $n_1$  [/fm$^3$]                & $n_2$  [/fm$^3$]               \\
\hline
mixB200 & 2.30E-01 & 1.29 \\  
mixB250 & 2.63E-01 & 1.51 \\ 
\hline
\end{tabular}
\end{minipage}
\end{table}
%
\subsection{Constructing method of equilibrium stellar configurations}\label{sec:basic}
Employing the constructed EOS with the phase transition, we construct the equilibrium 
stellar configurations. As mentioned in introduction, we pay attention to the 
general relativistic and toroidally magnetized stellar configurations. 
The basic equations 
and the numerical methods for the purpose are already given 
in \citet{kiuchi08b}. Hence, we only give a brief summary for later convenience. 

Assumptions to obtain the equilibrium models are summarized as follows 
; (1) Equilibrium models are stationary and axisymmetric. 
(2) The matter source is approximated by a perfect fluid with infinite conductivity. 
(3) There is no meridional flow of the matter. (4) The EOS for 
the matter is barotropic, which is satisfied as mentioned.
(5) Magnetic axis and rotation axis are aligned. 
 Because the circularity condition~\citep{wald84} holds under these assumptions, 
the metric can be written in the form
~\citep{komatsu89, cook92},
\begin{eqnarray}
ds^2 = - {\rm e}^{\gamma+\rho} dt^2 + {\rm e}^{2\alpha}( dr^2 + r^2 d\theta^2 ) \nonumber \\
+ {\rm e}^{\gamma-\rho}r^2\sin^2\theta( d\varphi -\omega dt )^2,\label{eq:metric}
\end{eqnarray}
where the metric potentials, $\gamma$, $\rho$, $\alpha$, and $\omega$, 
are functions of $r$ and $\theta$ only. We see that the non-zero component 
of Faraday tensor $F_{\mu\nu}$ in this coordinate is $F_{12}$. 
Integrability of the equation of motion of the matter requires,
\begin{eqnarray}
e^{\gamma -2\alpha} \sin{\theta} F_{12} = K(u);~~~u \equiv \rho_0  h e^{2\gamma} r^2 \sin^2{\theta} ,
\label{eq:I-con1}
\end{eqnarray}
where $K$ is an arbitrary function of $\rho_0 h e^{2\gamma}r^2\sin^2\theta$.  
The variables $\rho_0$ and $h$ represent 
the baryon density and relativistic specific enthalpy, respectively. 
Integrating the equation of motion of the matter, we arrive at the 
equation of hydrostatic equilibrium,

\begin{eqnarray}
\int^{P(\epsilon)}_{0} \frac{dP}{\epsilon +P} 
+ \frac{\rho +\gamma }{2}
+ \frac{1}{2} {\rm ln} (1-v^2) \nonumber \\
+ \frac{1}{4\pi} \int \frac{K(u)}{u}\frac{dK}{du}du = C,\label{eq:Ber}
\end{eqnarray}
where $v=(\Omega-\omega)r\sin\theta {\rm e}^{-\rho}$ with $\Omega$ being the 
angular velocity of the matter and $C$ is an integration constant. 
 Here, we assume the rigid rotation. The first term of equation (\ref{eq:Ber}) depends only on EOSs. Hence, we prepare the values of the integral with EOS tables, precisely. This preparation enable us to calculate equation (\ref{eq:Ber}) precisely, though our EOSs are quite different from polytrope models. To compute specific models of the magnetized stars, we need specify the function form of $K$, which determines the distribution of the magnetic fields. We take the following simple form,

\begin{eqnarray}
&& K(u) = b u^k,\label{eq:comp-mag1}
\end{eqnarray}
where $b$ and $k$ are constants. Regularity of toroidal magnetic field 
on the magnetic axis requires that $k \geq 1$. If $k \geq  1$, the magnetic 
fields vanish at the surface of the star. 
 In this study, we consider the $k=1$ case  
because in the general relativistic MHD simulation, \citet{kiuchi08c} have found 
that magnetic distribution with $k\ne1$ is unstable against axisymmetric perturbations.

To solve the master equations numerically, 
we employ the Cook-Shapiro-Teukolsky scheme~\citep{cook92} 
extended by \citet{kiuchi08b}, which does not care about the function 
form of EOS. Hence, it is straight forward to update our numerical code 
for incorporating the EOS with the phase transition.

After obtaining solutions, it is useful to compute global physical 
quantities characterizing 
the equilibrium configurations to clearly understand the properties of 
the sequences of the equilibrium models. In this paper, we compute the following 
quantities: the gravitational mass $M$, the baryon rest mass $M_0$, 
the total angular momentum $J$, the total rotational energy $T$, the total 
magnetic energy $H$, the magnetic flux $\Phi$, 
the gravitational energy $W$ and the mean deformation rate $\bar{e}$, whose 
definitions are explicitly given in \citet{kiuchi08b}. 
More explicitly, the mean deformation rate $\bar{e}$ is defined as 
\begin{eqnarray}
\bar{e}\equiv \frac{I_{zz}-I_{xx}}{I_{zz}},\label{eq:meandef}
\end{eqnarray}
where $I_{xx}=\pi\int\epsilon r^4 \sin\theta(1+\cos^2\theta)drd\theta$ 
and $I_{zz}=2\pi\int\epsilon r^4\sin^3\theta drd\theta$ with $\epsilon$ 
being the energy density of the matter. 
 Circumferential radius $R_{\rm cir}$ is defined as $R_{\rm cir}\equiv {\rm e}^{(\gamma-\rho)/2} r_e$ with $r_e$ being the coordinate radius at the stellar equatorial surface. 

We checked the convergence of the presented results 
by doubling the mesh numbers from the standard set of radial and angular direction 
mesh points of $400 \times 260$.
By checking the relativistic virial identities \citep{Bonazzola:1994} for all the models,
we confirm that the typical values are orders of magnitude $10^{-3}$, and become 
 $10^{-2}$ at worse ($10^{-4}$ at best).
These values, which is a measure of the numerical convergence, are almost same for the polytropic EOS case \citep{kiuchi08b}. 
In general, the convergence is known to become much worse for realistic EOSs, because 
their density and pressure profile are not smooth due to phase transitions. 
 In this respect, our numerical scheme works well.  
In Sec. \ref{ssec:rot}, we will discuss energy releases by the QCD phase transition and 
find that the values of our interest are $\sim 10^{-2}M_\odot$, for which the 
numerical accuracy above is certificated mostly. 
By doubling the mesh points, i.e., $800(r) \times 520(\theta)$, 
 we checked that the order of the magnitude of the released energy does not change.

In constructing one equilibrium sequence, we have three parameters to choose,
namely the central density $\rho_c$, the strength of the magnetic field 
parameter $b$, and the axis ratio $r_p/r_e$. Changing these parameters, we 
 seek solutions in as wide parameter range as possible to 
study the properties of the equilibrium sequences.
We need to calculate more than 50 models changing $\rho_c,\ b$, to follow 
one evolutionary sequence for a fixed baryon mass and         
magnetic flux. In addition, we change the initial angular momentum of each 
sequence by 10 models to model the rapidly rotating to the non-rotating case and also  3 different EOSs of mixB200, mixB250, and Shen are employed. 
This means that we have to construct at least 1500 models to explore properties 
of the magnetized compact objects with quark cores systematically.
In doing so, we use the Rosenbrock and Gram-Schmidt method, which is helpful 
 to obtain the convergence of the solutions efficiently.

\section{Numerical Results}\label{sec:result}
 First of all, we discuss how the EOS with the phase transition 
affects the equilibrium configurations in subsection \ref{sec:nonrot}, 
where we pay attention to non-rotating models. 
Since the magnetars and the high field neutron stars observed so far are all 
slow rotators, such non-rotating but highly magnetized 
static models could well be approximated to such stars. Moreover the static models merit 
that one can see purely magnetic effects on the equilibrium properties because there 
is no centrifugal force and all the stellar deformation is attributed 
to the magnetic stress.  Then in subsection \ref{ssec:rot}, we move on to discuss the 
 the releasable energy of the phase transition from the rotating and magnetized 
hadronic stars to the hybrid stars.

\subsection{Effect of the phase transition on the equilibrium configurations}\label{sec:nonrot}
\begin{figure*}
\includegraphics[width=17cm]{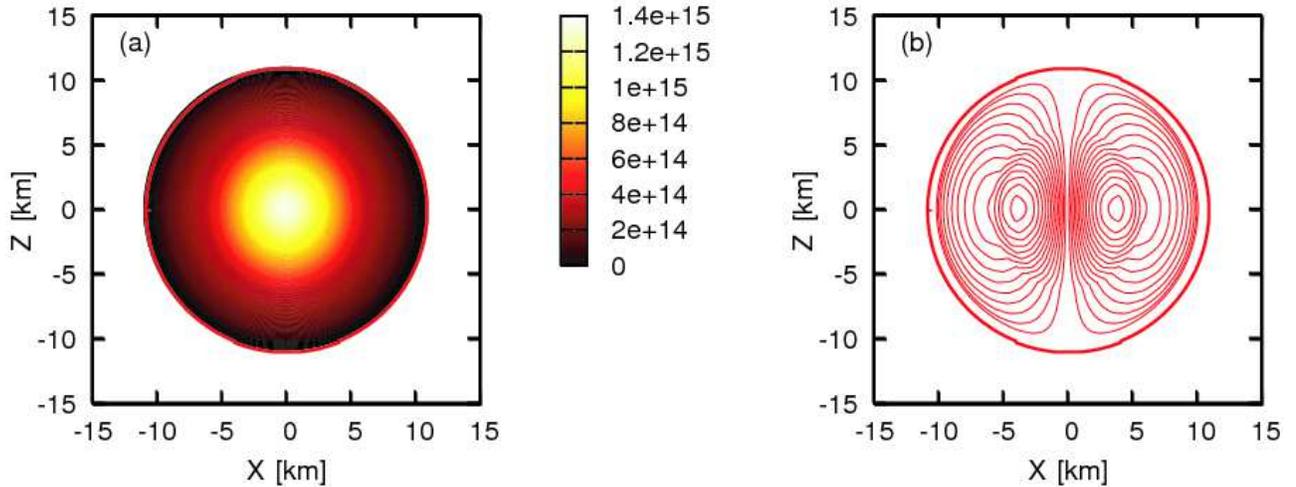} 
\caption{\label{fig:dis_mix200} Distribution of (a): baryon density [g/cm$^3$] and (b): magnetic field 
with "mixB200" EOS. Here, the central baryon density 
$\rho_{0,\rm max} =1.35 \times 10^{15}$ g cm$^{-3}$, the baryon mass $M_0=1.86 M_\odot$, 
the gravitational mass $M=1.69 M_\odot$,  the flux normalized by units of $10^{30}$~G cm$^2$, $\Phi_{30}=2.00$, the ratio of the magnetic energy to the gravitational 
energy, $H/|W|=1.65 \times 10^{-2}$, maximum magnetic fields, $B_{\rm max}=1.44 \times 10^{18}$ G. }
\end{figure*}

\begin{figure*}
\includegraphics[width=17cm]{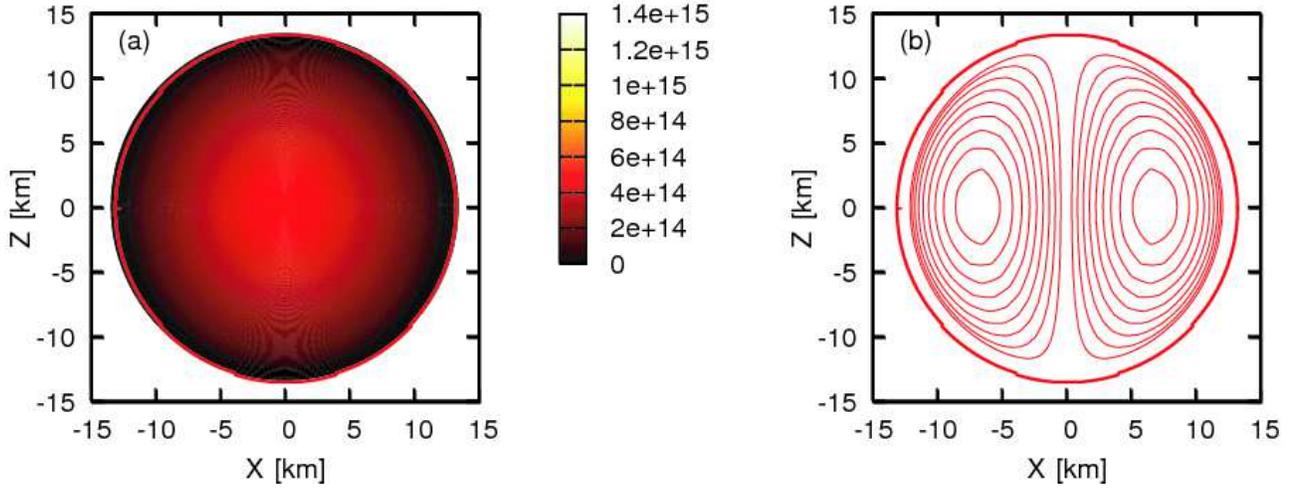} 
\caption{\label{fig:dis_shen} Same as Figure~\ref{fig:dis_shen} but for hadronic 
"Shen" EOS, $M_0=1.86 M_\odot$ star with the central baryon density $\rho_{0,\rm max} =4.86 \times 10^{14}$ g cm$^{-3}$, $M=1.71 M_\odot$, $H/|W|=2.25 \times 10^{-2}$, $B_{\rm max}=0.97\times 10^{18}$ G. }
\end{figure*}

Now we discuss the equilibrium configurations of the non-rotating and 
strongly magnetized compact stars with/without the quark cores.
 Figures \ref{fig:dis_mix200} and \ref{fig:dis_shen} are one example 
 for the hybrid star with mixB200 EOS and for the neutron star 
with the hadronic ``Shen'' EOS, 
showing the distributions of the baryon density (left panel) and 
the magnetic field (right panel) in the meridional planes, respectively. 
These two models have the 
same baryon mass of $1.86 M_\odot$, and the same magnetic flux 
of $2.00\times10^{30}{\rm G~cm^2}$, which mean that 
they are really highly magnetized models with the central magnetic fields of 
$\sim 10^{18}$ G. 

\begin{figure*}
\begin{center}
\hspace{-30mm}
\includegraphics[width=10cm]{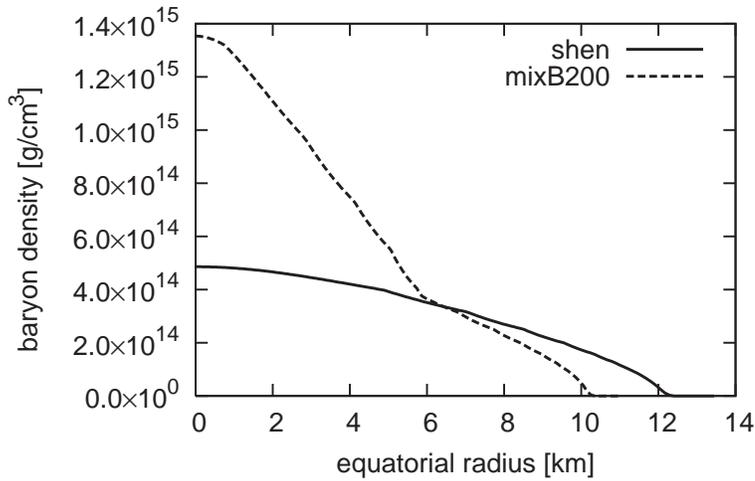} 
\caption{\label{fig:structure} Distribution of baryon density for Figure~\ref{fig:dis_mix200} and Figure~\ref{fig:dis_shen}. }
\end{center}
\end{figure*}

Comparing left panels of Figures \ref{fig:dis_mix200} and \ref{fig:dis_shen}, 
more concentration of the matter in the center 
is clearly seen for the hybrid star (Figure \ref{fig:dis_mix200}). 
The concentration is also clearly seen from Figure \ref{fig:structure}.
Inside the inner 6 km of the core of the hybrid star, the mixed phase appears, 
leading to the enhancement in the compression of the matter due to the softening of 
the EOS (e.g., Figures 1 and 2).

From right panel of Figure \ref{fig:dis_shen}, the strong toroidal field 
lines are seen to behave like a rubber belt, wrapping around the 
waist of the neutron star. 
It is found that the magnetic fields frozen-in to the matter, are also 
compressed by the presence of the quark phase for the hybrid star (right panel 
of Figure \ref{fig:dis_mix200}). In fact, the pinching of the field lines in the panel corresponds to 
the surface of the 
quark core, $\sim$ 6km in radius. These qualitative features are also 
true for the other equilibrium models.

In addition, a general trend in the equilibrium configurations for the hybrid stars is 
that they are more compact than the neutron star, due to the softness of 
the EOS. Given the same stellar baryon mass ($M_0$), 
the gravitational mass is smaller up to $\sim 0.01 M_\odot $ 
than for the neutron stars, reflecting their smaller energy of the quark matter 
(right panel of Figure \ref{fig:eos1}). 

Given the fixed magnetic flux, we can construct equilibrium sequences by changing the 
central density. To characterize the features of the hybrid stars, 
we pay attention to the model with the maximum mass ($d M/ d\rho_{0,c} = 0$) 
along each sequence, which we call as the maximum mass sequence for convenience.
In table 2, the important physical quantities are summarized for the maximum mass sequences 
 with different magnetic fluxes. It is here 
noted that the maximum value of the magnetic flux
 in the table ($\Phi = 2.5 \times 10^{30} {\rm G~cm^2}$) corresponds to the
 non-convergence limit, beyond which any solutions cannot converge with the present 
numerical scheme \citep{kiuchi08b}. Albeit with such limitations, 
the field strength is already enough high to affect the configurations 
and we can well study the magnetic effects on them.

Since the compression of the matter is more enhanced for the smaller bag constant, 
 the hybrid stars with the smaller bag constant become 
more compact (see $R_{\rm cir}$ Table 2).
 The maximum magnetic fields are found to become up to about 30 \% larger 
for the hybrid stars than for the neutron stars (compare $B_{\rm max}$ for 
mixB200 with for Shen). 
 It is also found that the hybrid stars with the smaller bag 
constant become more prolate (smaller values of $\bar{e}$) and also 
 their maximum masses become smaller up to $\sim 10\%$ than those 
for the larger bag constant models. 
All these features are helpful to understand the properties of the evolution tracks of the hybrid stars, which we discuss from the next section.
 
\begin{table*}
\centering
\begin{minipage}{140mm}
\caption{\label{tab:nonrot-ADMmax}
Global physical quantities for the maximum gravitational mass models of the constant 
magnetic flux sequences of the non-rotating stars. The order of EOSs from top to down 
indicates for their hardness from the hadronic "Shen EOS" to the softest "mixB200" EOS.
}
\begin{tabular}{cccccccc}
\hline\hline
$\Phi[10^{30}{\rm G~cm^2}]$          &
$\rho_{0,c}[10^{15}{\rm g}/{\rm cm}^3]$ &
$M[M_{\rm \odot}]$     &
$M_0[M_{\rm \odot}]$                &
$R_{\rm cir}[{\rm km}]$          &
$B_{\rm max}[10^{18}{\rm G}]$    &
$H/|W|$                             &
$\bar{e}$                          \\
\hline
\multicolumn{8}{c}{Shen}\\
\hline
  0.00E+00&1.16E+00&2.46E+00&2.86E+00&1.35E+01&0.00E+00&0.00E+00& 0.00E+00\\  
  1.50E+00&1.01E+00&2.45E+00&2.83E+00&1.39E+01&7.34E-01&3.81E-03&-7.40E-03\\  
  2.00E+00&1.05E+00&2.45E+00&2.84E+00&1.39E+01&9.83E-01&6.59E-03&-1.30E-02\\ 
  2.50E+00&1.17E+00&2.46E+00&2.85E+00&1.36E+01&1.26E+00&9.48E-03&-1.86E-02\\ 
\hline
\multicolumn{8}{c}{mixB250}\\
\hline
  0.00E+00&1.18E+00&1.83E+00&2.03E+00&1.47E+01&0.00E+00&0.00E+00&  0.00E+00\\ 
  1.50E+00&1.17E+00&1.83E+00&2.02E+00&1.40E+01&9.38E-01&8.28E-03&-1.97E-02\\ 
  2.00E+00&1.18E+00&1.83E+00&2.02E+00&1.41E+01&1.23E+00&1.45E-02&-3.50E-02\\ 
  2.50E+00&1.64E+00&1.82E+00&2.00E+00&1.33E+01&1.76E+00&2.02E-02&-4.63E-02\\ 
\hline
\multicolumn{8}{c}{mixB200}\\
\hline
  0.00E+00&1.44E+00&1.70E+00&1.88E+00&1.30E+01&0.00E+00&0.00E+00&  0.00E+00\\ 
  1.50E+00&1.29E+00&1.69E+00&1.86E+00&1.34E+01&1.08E+00&9.62E-03&-2.25E-02\\ 
  2.00E+00&1.35E+00&1.69E+00&1.86E+00&1.34E+01&1.44E+00&1.65E-02&-3.92E-02\\ 
  2.50E+00&1.65E+00&1.69E+00&1.86E+00&1.31E+01&1.86E+00&2.39E-02&-5.59E-02\\ 
\hline
\end{tabular}
\end{minipage}
\end{table*}

\subsection{An evolutionary track to a hybrid star}\label{ssec:rot}
Based on the equilibrium configurations mentioned above, 
we now move on to discuss an evolutionary 
path to the formation of a hybrid star due to the spin-down of magnetized and 
 rotating neutron stars. 

\begin{figure*}
\begin{center}
\vspace*{-110mm}
\includegraphics[width=16cm]{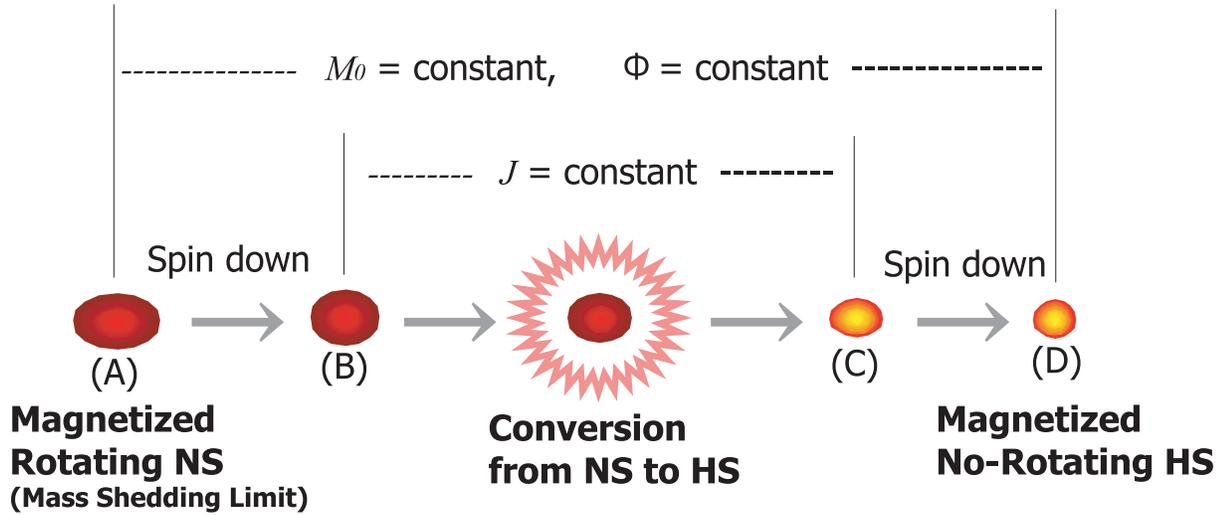} 
\caption{\label{fig:senario} Schematic drawing, showing our scenario 
of conversion from neutron stars~(NSs) to hybrid stars~(HSs). The baryon rest mass and the magnetic flux are assumed to conserve for modeling
 the isolated neutron stars that are adiabatically losing angular momentum via the gravitational radiation and/or magnetic breaking. We furthermore assume the conservation of the angular momentum at the conversion.}
\end{center}
\end{figure*}

The evolution scenario we have in mind is illustrated in Figure \ref{fig:senario}.
Let us consider evolution of a single protoneutron star, left after core-collapse 
supernova explosion, which rotates with the mass-shedding limit ((A) in Figure 
\ref{fig:senario}). The maximum magnetic fields 
deep inside the core are taken to be $\sim 10^{18}$ G, which could be sustained 
 due to $\alpha - \Omega$ dynamos in such a rapidly rotating neutron 
star \citep{Thompson:1993hn,Thompson:1996pe}.  It should be noted that 
 the possibility of such ultra-magnetic fields has not been rejected so far, because
what we can learn from the observations of magnetars 
by their periods and spin-down rates 
is only their surface fields ($\sim 10^{15}$ G).
 During their evolution, the baryon mass
 and the magnetic field strength are assumed to be constant for simplicity.
Such models may model the evolution of the isolated compact stars,   
losing angular momentum via the gravitational radiation and/or magnetic breaking 
(from (A) to (B)).
The phase transition to the hybrid stars is expected to take place during the evolution
(shown as (B) to (C) in Figure~\ref{fig:senario}).
At the moment, the baryon rest masses and the angular momenta are assumed to be conserved, which may be justified because
 the timescale of the conversion, albeit uncertain of 0.1-100~s, are too short 
compared to typical evolutionary timescale of compact stars (more than $10^3$ years)
 \citep{olesen91}. 
After the transition, the newly born hybrid star evolves, again losing the angular 
momentum, to settle down to the magnetized and no-rotating hybrid star finally 
(see (C) to (D)).

\begin{figure*}
\includegraphics[width=8cm]{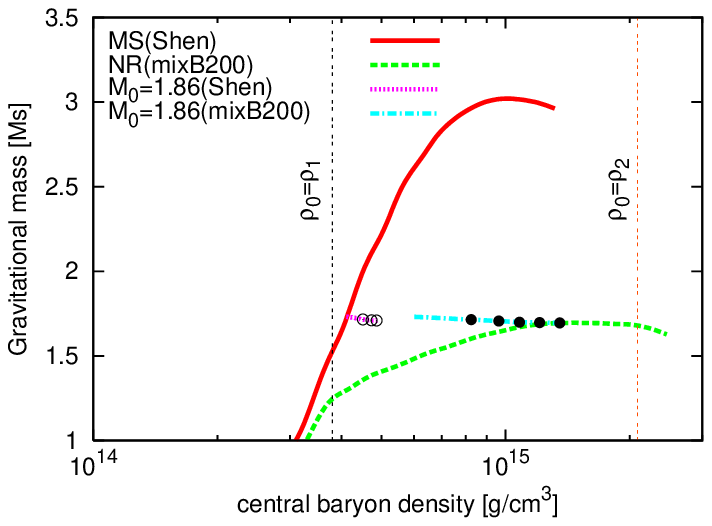} 
\includegraphics[width=8cm]{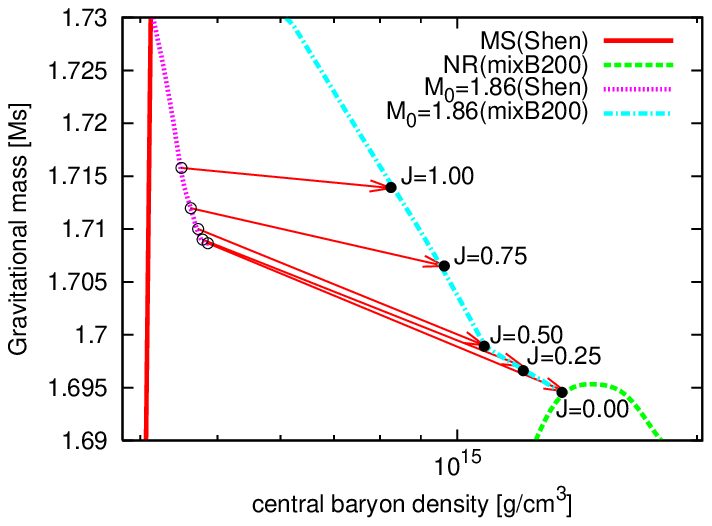} 
\caption{\label{fig:Rdis_mix200} Gravitational masses versus central effective baryon densities at the constant magnetic flux, $\Phi_{30}=2.00$. The right panel is magnified figure of the left panel. The solid line of MS(Shen) is for mass shedding limits of neutron stars. The dashed line of NR(mixB200) is for no-rotating hybrid stars with "mixB200" EOS. The line of '$M_0=1.86$~(Shen)' is one example in all baryon mass constant lines for neutron stars. '$M_0=1.86$~(mixB200)' is same as '$M_0=1.86$~(Shen)', but for hybrid stars. Conversions from neutron stars to hybrid stars are shown as the vectors from circle~($\circ$) to filled circle~($\bullet$). Here, $J$ is the angular momentum in unit of $10^{49}$ g cm$^2$ s$^{-1}$. The lines of $\rho_1$ and $\rho_2$ are critical densities of phase equilibrium for hadron phase and quark phase shown in Figure~1. and Table~1. }
\end{figure*}

Now using the left panel of Figure \ref{fig:Rdis_mix200}, we proceed to discuss 
quantitatively the evolution tracks mentioned above. 
The red solid  the baryon mass '$M_0=1.86 M_\odot$' rotating 
with the mass shedding limit (corresponding to (A) in Figure \ref{fig:senario}). 
It is noted that the choice of the baryon mass ($M_0=1.86 M_\odot$) 
is determined by the maximum mass of the non-rotating hybrid star with the mixB200 EOS 
as discussed in subsection \ref{sec:nonrot} and also that
 this baryon mass constant sequence has the largest releasable energy at the transition 
among all the models, as we show this point later. 
There are many possible paths to convert to hybrid stars with different masses. However, we focused on the maximum energy release with QCD phase transition in this paper.
It is possible to select a single equilibrium 
sequence by keeping the baryon rest mass and the magnetic flux constant simultaneously. 
Thus the evolution of the protoneutron star can be described along the line 
(pink dotted line), going right from one point on the red solid curve.
 As mentioned, our models assume that all equilibrium sequences begin at their 
mass-shedding limits and continue to non-rotating equilibrium hybrid stars, 
at which the sequences end.
The final fate is shown by the green dashed line in Figure \ref{fig:Rdis_mix200}, which is the 
 non-rotating hybrid star with the same baryon mass of '$M_0=1.86 M_\odot$', here 
for the mixB200 EOS (corresponding (D) in Figure \ref{fig:senario}). 

Between $\rho_1$ and $\rho_2$ (the vertical lines in Figure \ref{fig:Rdis_mix200}), the phase transition 
should take place, 
producing the jump to the sequence of hybrid stars (from (B) to (C) in Figure \ref{fig:senario}). 
To see clearly, the transition is shown as 
the vectors from circle~($\circ$) to filled circle($\bullet$) in the right panel 
of Figure \ref{fig:Rdis_mix200}, which is just the magnification of the left panel 
near the phase transition. Here, '$J=1.00$'~('$J=0.70$',~'$J=0.50$',~'$J=0.25$') 
denotes the angular momentum in unit of $10^{49}$ g cm$^2$ s$^{-1}$.
We estimate the energy release at the phase transition simply by 
the difference in the mass-energies 
between the neutron star and the hybrid star,
\begin{eqnarray}
\Delta M = M_{NS}(M_0, \Phi, J) - M_{HS}(M_0, \Phi, J), 
\end{eqnarray}
with conserving $\Phi$ and $J$ through the transition.

\begin{figure*}
\hspace*{-5mm}
\includegraphics[width=8cm]{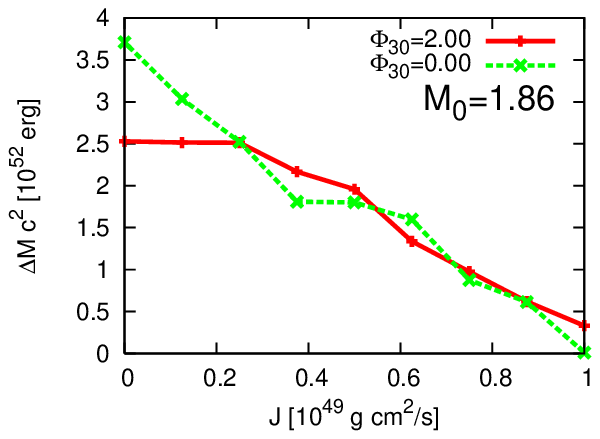} \\
\includegraphics[width=8cm]{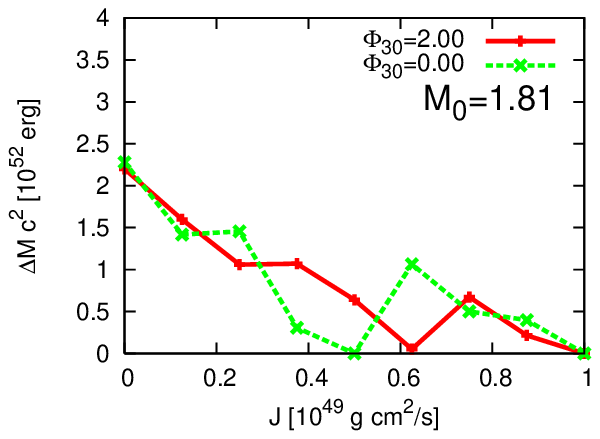} \\
\includegraphics[width=8cm]{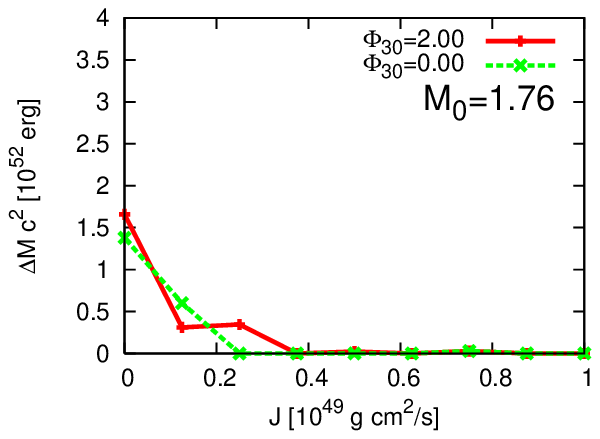}
\caption{\label{deltaM} Energy release ($\Delta M$) 
as a function of the angular momentum ($J$) at the 
moment of the phase transition for magnetized ($\Phi_{30}=2.00$) 
and non-magnetized models ($\Phi_{30}=0 $) for $M_0=1.86 M_\odot$~(the upper panel), $M_0=1.81 M_\odot$~(the middle panel), and $M_0=1.76 M_\odot$~(the lower panel). 
 Note here that these are the case for the mixB200 EOS. 
For the mixB250 EOS,  exact values of $\Delta M$ 
cannot be estimated due to the limitation of the present scheme. Unfortunately
 it is also seen here, for example, for the case of smaller baryon mass 
 for $J \ga 0.4 \times 10^{49} 
[{\rm g}~{\rm cm}^2~{\rm  s}^{-1}]$ (bottom)(see text for more details).}
\end{figure*}

In Figure \ref{deltaM}, we show the releasable energy of $\Delta M$ 
 as a function of the angular momentum for the mixB200 EOS. 
Top panel is for the baryon mass of 
 $M_0=1.86 M_\odot$, which is a special evolutionary 
track in the sense that the baryon mass is the maximum 
among non-rotating models, so that the releasable energy is largest among the computed 
models.
 Also for the other baryon mass models (middle and bottom panels), 
it is seen that the energy release becomes smaller for the larger 
angular momentum. This is simply because stronger centrifugal forces prevent 
the compression (see the central density ($\rho_c$)
 in Tables \ref{tab:Mb-const-200} and \ref{tab:Mb-const-2}), making the 
quark cores smaller, which weakens the effect of the phase transition. 
In all the computed models, the maximum energy release is found to be 
$ \la 0.01 M_{\rm \odot}$. 
This energy is equivalent to $ \la 2 \times 10^{52}$ erg, which is truly 
larger compared with the energies of supernovae ($\sim 10^{51}$ erg), however much 
smaller than those previously estimated ($\sim 10^{53}$ erg) in which 
the conversion to the strange quark star is assumed 
\citep{gourgoulhon99,yasutake05, zdunik07}. 
We note that $\Delta M$ becomes 
less than $\Delta M \la 10^{-3} M_{\odot}$ for the larger bag constant models of 
 mixB250.  This is because the large bag constant suppresses the conversion 
from hadron matter to quark matter as already mentioned. Unfortunately,
   exact values of $\Delta M$ for the mixB250 models
 cannot be found by the present scheme because $\Delta M$ becomes more than $3$ orders-of-magnitude smaller than $M,$ while $M$ can be estimated at most with the numerical 
accuracy of $10^{-2}$ for this case.
This is also the case for small baryon mass cases (see the bottom panel of Figure~8).
For getting more precise solutions, we may need to employ 
the so-called spectral method. Although the LORENE code \citep{bona} employs the method, 
 it cannot treat the magnetic fields yet. The implementation of the method is 
 still a major undertaking, which we pose as a next task of this paper.

\begin{figure*}
\includegraphics[width=10cm]{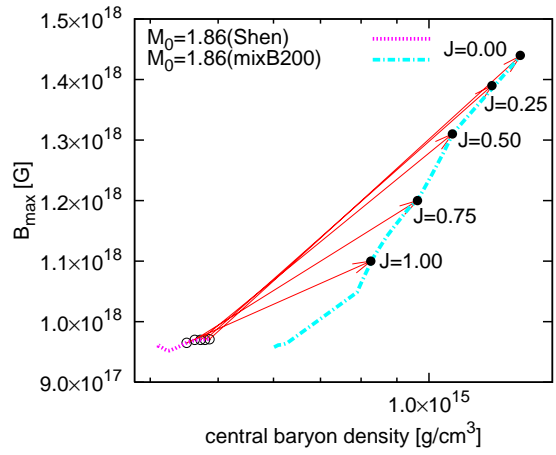}
\caption{\label{fig:b_up} Central baryon density versus maximum magnetic field for same stars of Figure~\ref{fig:Rdis_mix200}.}
\end{figure*}

\begin{figure*}
\begin{center}
\includegraphics[width=8cm]{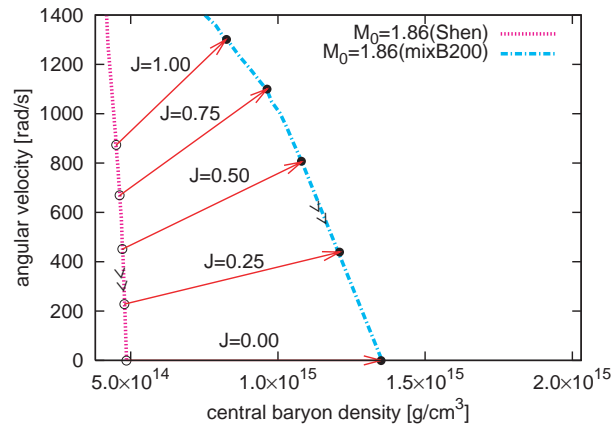} 
\caption{\label{fig:omega} Central baryon density versus angular velocities for same stars of Figure~\ref{fig:Rdis_mix200}. Angular velocities are monotonically decreased without the conversion, shown as '$>>$'.}
\end{center}
\end{figure*}

Other numerical values characterizing the phase transition are given 
in Tables \ref{tab:Mb-const-200} and \ref{tab:Mb-const-2}. 
 Table~\ref{tab:Mb-const-2} is for the non-magnetized models, which is given 
for comparison.
For models with smaller angular momentum ($J$ in the tables), it can be readily seen 
that the gravitational masses ($M$) and the circumferential radii ($R_{\rm cir}$) 
become smaller, while the central baryon densities ($\rho_{0,c}$) and the maximum strength of 
the magnetic fields ($B_{\rm max}$) become larger, all simply due to the smaller
 centrifugal forces. If the newly-born neutron star evolve to the non-rotating 
neutron star without experiencing the conversion, the increase of the central 
density is less than $\sim 20 \%$ (compare $\rho_{0,c}$ of 
"Shen" at the mass shedding limit with the one at no-rotating limit 
in Table \ref{tab:Mb-const-200}).
On the other hand, the maximum density is found to increase about several times 
larger by the phase transition 
(compare $\rho_{0,c}$ between "Shen" with "mixB200" in Table \ref{tab:Mb-const-200}). 
Frozen-in to the matter, the maximum magnetic field increases up to $50\%$ while 
the change is an order of unit percent in absence of the conversion 
(see Figure \ref{fig:b_up}). 
Due to the conservation of 
angular momentum, the angular velocity increases 
about $\Delta \Omega / \Omega \sim 30 \%$ after the conversion 
(see Figure \ref{fig:omega} and Tables \ref{tab:Mb-const-200}). 
Also in this case, it is noted that 
the spin-up is suppressed for the larger bag constant models (see Table 3).
We speculate that such spin-up, much larger than the canonical glitches of magnetars 
($\Delta \Omega / \Omega \sim 10^{-7}$, \citet{woods}), could be new observable 
signatures, 
marking the occurrence of the phase transition.

\begin{table*}
\centering
\begin{minipage}{180mm}
\caption{\label{tab:Mb-const-200}
Global physical quantities for the equilibrium sequences of the rotating 
stars with $M_0 = 1.86 M_\odot$, $\Phi_{30}=2.00$. 
}
\begin{tabular}{ccccccccc}
\hline\hline
EOS &
$\rho_{0,c}[10^{15}{\rm g}/{\rm cm}^3]$ &
$M[M_{\rm \odot}]$                   &
$R_{\rm cir}[{\rm km}]$              &
$\Omega$ [10$^3$ rad/s]              & 
$T/|W|$                                   &
$H/|W|$                                  &
$\bar{e}$                               &
$B_{\rm max}[10^{18}{\rm G}]$          \\
\hline
Shen(MS)    &4.11E-01&1.73E+00&2.22E+01&1.59E+00&9.41E-02&2.89E-02&2.00E-01&9.61E-01\\
\hline
\multicolumn{7}{c}{ $J$ = 1.00 $\times$ 10$^{49}$ g cm$^2$/s }\\
\hline
Shen     & 4.51E-01&1.72E+00&1.70E+01&8.73E-01&3.69E-02&2.47E-02& 5.40E-02&9.65E-01\\
mixB250  & 5.11E-01&1.72E+00&1.69E+01&9.31E-01&3.65E-02&2.42E-02&5.40E-02&9.68E-01\\
mixB200  & 8.26E-01&1.71E+00&1.61E+01&1.30E+00&3.78E-02&2.11E-02&  7.01E-02&1.10E+00\\
\hline
\multicolumn{7}{c}{ $J$ = 0.75 $\times$ 10$^{49}$ g cm$^2$/s }\\
\hline
Shen&4.63E-01&1.71+00&1.64E+01&6.69E-01&2.15E-02&2.37E-02&7.26E-03&9.70E-01\\
mixB250&5.56E+00&1.71E+00&1.63E+01&7.40E-01&2.13E-02&2.33E-02&8.32E-03&9.74E-01\\
mixB200&9.63E-01&1.71E+00&1.51E+01&1.10E+00&2.24E-02&1.93E-02&2.74E-02&1.20E+00\\
\hline
\multicolumn{7}{c}{ $J$ = 0.50 $\times$ 10$^{49}$ g cm$^2$/s }\\
\hline
Shen    &4.72E-01&1.71E+00&1.60E+01&4.54E-01&9.83E-03&2.31E-02&-3.05E-02&9.70E-01\\
mixB250 &5.93E-01&1.71E+00&1.59E+01&5.16E-01&9.77E-03&2.24E-02&-2.82E-02&9.75E-01\\
mixB200 & 1.08E+00&1.70E+00&1.43E+01&8.07E-01&1.04E-02&1.81E-02&-8.49E-03&1.31E+00\\
\hline
\multicolumn{7}{c}{ $J$ = 0.25 $\times$ 10$^{49}$ g cm$^2$/s }\\
\hline
Shen&4.79E-01&1.71E+00&1.58E+01&2.28E-01&4.53E-03&2.28E-02&-4.82E-02&9.70E-01\\
mixB250&6.15E-01&1.71E+00&1.57E+01&2.60E-01&4.43E-03&2.20E-02&-4.57E-02&9.75E-01\\
mixB200&1.21E+00&1.69E+00&1.42E+01&4.39E-01&2.82E-03&1.71E-02&-3.15E-02&1.39E+00\\
\hline
\multicolumn{7}{c}{ no rotation }\\
\hline
Shen     &4.86E-01&1.71E+00&1.57E+01&0.00E+00&0.00E+00&2.25E-02&-6.35E-02&9.72E-01\\
mixB250  &6.36E-01&1.71E+00&1.55E+01&0.00E+00&0.00E+00&2.16E-02&-6.03E-02&9.75E-01\\
mixB200  & 1.35E+00&1.69E+00&1.34E+01&0.00E+00&0.00E+00&1.65E-02&-3.92E-02&1.44E+00\\
\hline
\hline
\end{tabular}
\end{minipage}
\end{table*}

\begin{table*}
\centering
\begin{minipage}{180mm}
\caption{\label{tab:Mb-const-2}
Same as Table~\ref{tab:Mb-const-200} but for $\Phi_{30}=0$. }
\begin{tabular}{ccccccccc}
\hline\hline
EOS &
$\rho_{0,c}[10^{15}{\rm g}/{\rm cm}^3]$ &
$M[M_{\rm \odot}]$                       &
$R_{\rm cir} [{\rm{km}}]$              &
$\Omega$ [10$^3$ rad/s]                  & 
$T/|W|$                                       &
$\bar{e}$                                   \\
\hline
Shen(MS)     &3.81E-01&1.74E+00&2.10E+01&1.45E+00&1.12E-01&2.79E-01\\
\hline
\multicolumn{7}{c}{ $J$ = 1.00 $\times$ 10$^{49}$ g cm$^2$/s }\\
\hline
Shen      &4.40E-01&1.72E+00&1.66E+01&8.08E-01&3.58E-02&1.07E-01\\
mixB250   &4.62E-01&1.72E+00&1.66E+01&8.30E-01&3.54E-02&1.06E-01\\
mixB200   &7.80E-01&1.71E+00&1.58E+01&1.18E+00&3.72E-02&1.11E-01\\
\hline
\multicolumn{7}{c}{ $J$ = 0.75 $\times$ 10$^{49}$ g cm$^2$/s }\\
\hline
Shen    &4.51E-01&1.71E+00&1.61E+01&6.23E-01&2.09E-02&6.49E-02\\
mixB250 &5.16E-01&1.71E+00&1.61E+01&6.69E-01&2.07E-02&6.43E-02\\
mixB200 &9.02E-01&1.70E+00&1.49E+01&1.01E+00&2.18E-02&6.90E-02\\
\hline
\multicolumn{7}{c}{ $J$ = 0.50 $\times$ 10$^{49}$ g cm$^2$/s }\\
\hline
Shen      &4.61E-01&1.71E+00&1.58E+01&4.25E-01&9.54E-03&3.06E-02\\
mixB250   &5.51E-01&1.71E+00&1.57E+01&4.69E-01&9.49E-03&3.04E-02\\
mixB200   &1.03E+00&1.70E+00&1.42E+01&7.51E-01&1.02E-02&3.33E-02\\
\hline
\multicolumn{7}{c}{ $J$ = 0.25 $\times$ 10$^{49}$ g cm$^2$/s }\\
\hline
Shen    &4.67E-01&1.71E+00&1.56E+01&2.15E-01&2.54E-03&8.48E-03\\
mixB250 &5.74E-01&1.71E+00&1.55E+01&2.42E-01&2.52E-03&8.46E-03\\
mixB200 &1.14E+00&1.69E+00&1.37E+01&4.08E-01&2.65E-03&9.10E-03\\
\hline
\multicolumn{7}{c}{ no rotation }\\
\hline
Shen      &4.75E-01&1.71E+00&1.55E+01&0.00E+00&0.00E+00&0.00E+00\\
mixB250   &5.82E-01&1.71E+00&1.54E+01&0.00E+00&0.00E+00&0.00E+00\\
mixB200   &1.19E+00&1.69E+00&1.35E+01&0.00E+00&0.00E+00&0.00E+00\\
\hline
\end{tabular}
\end{minipage}
\end{table*}

\section{Summary and Discussion}\label{sec:summary}
In this study, we investigated structures of general relativistic hybrid stars
 containing super-strong magnetic fields. Pushed by the results of 
recent stellar evolution calculations and the outcomes of recent 
MHD simulations of core-collapse supernovae,
 we treated the toroidal fields only.
 Using the bulk Gibbs construction, we modeled
the EOS with a first order transition 
by bridging the MIT bag model for the description of quark matter 
and the nuclear EOS by Shen et al. We found that the presence of the quark phase 
can affect the distribution 
of the magnetic fields inside the hybrid stars, leading 
to the enhancement of the field strength about $30$ \% than for the normal neutron 
stars. Using the equilibrium configurations, we explored the possible evolutionary 
paths to the formation of hybrid stars due to the spin-down of magnetized and rotating 
neutron stars. For simplicity,  the total baryon mass and magnetic flux are 
taken to be conserved during the evolution but that the angular momenta are lost 
gravitational waves and/or magnetic breaking. 
We found that the energy release by the conversion to hybrid stars 
is typically $\la 10^{52}~\rm erg $, smaller than previously estimated 
for the conversion to the strange quark stars. 

 In association with the vast energy release, it is natural to expect the 
emissions of gravitational waves.
Amplitude of the gravitational waves associated with the phase transition, $h$, 
is given as follows 
\citep{pacheco98},
\begin{eqnarray}
h \sim 2 \left( \frac{G \alpha E_{\rm tot}}{c^3 \tau} \right)^{1/2} \frac{1}{r \omega}.
\end{eqnarray}
 Here, $E_{\rm tot}$, $\alpha$, $\tau$, $r$, and $\omega$ are energy release, 
 fraction of the energy release emitted in form of the gravitational waves, 
 damping time scale of 
the gravitational wave, distance to the source, and angular velocity, respectively. 
To set an absolute upper bound, we could choose $ E_{\rm tot} 
\sim 10^{52}$ erg and  $\omega \sim 100$ rad s$^{-1}$ 
according to Table~\ref{tab:Mb-const-200}.
 We assume that the source is in our galactic center of $r \sim 10$ kpc and 
that the damping timescale and $\alpha$ is $\sim 100$ ms and $10^{-4}$, respectively
 inferred from \citet{abdikamalov08}. The resulting amplitudes become as high 
as $h \sim 10^{-18 \sim -19}$ peaking at $\sim$ kHz, which are surely within 
the detection limits for the laser interferometers on line such as LIGO, VIRGO, GEO600, and TAMA300. Considering the optimal sensitivity of the interferometers of $h \sim 10^{-21}$ at $\sim$ 1 KHz, such signals, far stronger than the ones from
 canonical core-collapse supernovae  (e.g., \citet{kotake09a,kotake09b,murphy}), 
are possibly visible for a source out to Megaparsec distances. 
In combination with the signatures of the spin-up discussed 
before, we speculate that the waveforms of such strong gravitational waves could provide us hints of properties of QCD phase transition.
 To find the accurate waveforms, general relativistic simulations are 
indispensable (Kiuchi et al. in preparation). Moreover accurate predictions for 
 neutrino emissions \citep{gentile,nakazato08,sagert08} and their detectability are 
remained to be studied (e.g., \citet{kawagoe}). This paper should help to construct 
 the initial conditions for such studies.

\citet{alcock86} claimed that the conversion from the hadronic 
to the mixed phase occurs instantaneously in the weak interaction timescales~(such as $ d+u \leftrightarrow  s+u$) when the central density exceeds the critical value of $\rho_1$ in 
Figures 1 and 2. If so, the newly-born neutron stars may convert to the hybrid stars
 in the postbounce phase, before they evolve from A to B 
illustrated in Figure \ref{fig:senario}. 
In this case, the released energy may be used to power original 
 core-collapse supernovae or more energetic supernovae such as hypernovae.
Although the formation paths to hybrid stars are different from the ones discussed in this paper, the estimation method of the 
energy release here are also applicable if we implement the EOS
 at finite temperature and high lepton fraction \citep{yasutake09}, which we 
 plan to study as a sequel to this paper. 

It has been pointed out that the super-strong magnetic field more than $10^{18}$~G 
can affect the stiffness of the EOS \citep{broderick00}. 
As mentioned, the maximum magnetic field increases about $\sim 30 \%$ 
by the phase transition, while it increases only unit percent if the compact stars 
evolve to the non-rotating ones in absence of the transition.
This suggests that the magnetic effects should be seriously taken into account to 
our models treating the phase transition. Especially for the quark phase, \citet{fukushima07, noronha07} 
have recently found that 
the energy gaps of magnetic color-flavor-locked phase are oscillating 
functions of the magnetic field. Such effects are remained to be studied.

%
%
We should comment about the very high values of the magnetic fields, larger than $10^{18}$G. 
\citet{ruderman00} suggested that such high magnetic fields can be brought to the stellar surface by buoyancy forces. 
According to their result, we estimate the buoyancy time scale in our models. 
The buoyancy force can be expressed as 
\begin{eqnarray}
F_b \sim \frac{B_\phi^2}{8\pi c_s^2} g, 
\label{eq:fb}
\end{eqnarray}
where $c_s$ and $g$ are the sound speed and the gravitational acceleration, respectively. 
Putting typical values inside compact objects into the equation, 
we roughly estimate the time scale $\tau$ as 
\begin{eqnarray}
\tau \sim 10^{-3}-10^{-4} {\rm s} \left(\frac{\rho}{10^{15} {\rm g/cm^3}}\right)^{1/2}
\left(\frac{R}{10^6 {\rm cm}}\right)^{1/2}
\left(\frac{c_s}{10^{10}{\rm cm/s}}\right) 
\left(\frac{B_\phi}{10^{18} {\rm G}}\right)^{-1}
\left(\frac{g}{10^{14} {\rm cm /s^{2}}}\right)^{-1/2},
\end{eqnarray}
where $R$ is the typical size of compact stars.
Clearly such magnetic fields are not stable against the buoyancy forces in stellar 
evolution time scale $\sim 10^3$yr. Based on the MHD simulations in full general 
relativity, 
\citet{kiuchi08c} recently showed that
 the toroidal configurations of neutron stars is really dynamically unstable 
due to the buoyant instability. But more important finding relevant to this study, 
is that the toroidal magnetic fields settle down to a new equilibrium state with 
the circular motions in the meridian plane. In the new equilibrium configurations,  
the toroidal fields almost equivalent to the strength before the onset of the 
instability, are expected to be much stronger that the poloidal ones. This suggests that 
our models presented here, albeit unstable to the buoyant instability, 
could be helpful to understand equilibrium configurations of 
  magnetized hybrid stars. Such stability analysis 
is an important issue yet to be studied.

Our treatments of the phase transition should be more sophisticated. We plan to 
employ the so-called NJL model (e.g., \citet{yasutake09}) 
instead of the simple MIT bag model. 
Moreover we shall take into account finite size effects \citep{maruyama} 
such as quark-nuclear pasta structures.
We considered cold compact objects, namely with 
zero-temperature and zero-neutrino fraction. We plan to extend this study to the 
finite temperature with the non-zero neutrino fraction, which should be useful 
for studying earlier evolutions of compact stars soon after their formation. 

The evolution sequence we explored is a very simplified one.
For more realistic estimations, one needs the two 
dimensional evolutionary calculation in full general relativistic 
framework, not an easy job, due to the treatment of convection, combustion,
 cooling and/or heating processes, nucleosynthesis on the surface, and so forth. 
Applying the method recently reported by \cite{pons}, we think that 
we could study 
the magneto-thermal evolution of neutron stars to the hybrid stars.
By doing so, we hope to investigate the peculiar properties in the light curves, 
which has been pointed out to be another observationally visible 
\citep{page00,blaschke00,blaschke01,grigorian05}. All of such studies should be
 indispensable to pave the way for the understanding of the 
long-veiled phase-transition physics from the astrophysical phenomena.

\section*{Acknowledgments}
We are grateful to Shijun Yoshida, Y. Eriguchi, and M. Hashimoto, 
for informative discussions. N.Y. thanks to Y. Sekiguchi for stimulating 
 discussions. K.K. is grateful to K. Sato and S. Yamada for continuing encouragements. 
Numerical computations were in part carried on XT4 general common use computer system at the center for Computational Astrophysics, 
CfCA, the National Astronomical Observatory of Japan.  This
study was supported in part by the Grants-in-Aid for the Scientific Research 
from the Ministry of Education, Science and Culture of Japan (No. 
S19104006, 20740150, 21105512).

\clearpage

%

\clearpage

\bsp

\label{lastpage}

\end{document}